\documentclass[preprint]{elsarticle}

\usepackage{hyperref}

\journal{NeuroImage}

\usepackage{float}

\usepackage{amsmath}

\begin{document}

\begin{frontmatter}

\title{Independent components of human brain morphology.}

\author[ICOS,IONNCL,IONUCL]{Yujiang Wang\corref{mycorrespondingauthor}}
\author[Tuebingen]{Tobias Ludwig}
\author[ICOS,IONNCL]{Bethany Little}
\author[ICOS,IONNCL]{Joe H Necus}
\author[IONUCL,QU,Chalfont]{Gavin Winston}
\author[IONUCL,CMIC,Dep,Chalfont]{Sjoerd B Vos}
\author[IONUCL]{Jane de Tisi}
\author[IONUCL,Chalfont]{John S Duncan}
\author[ICOS,IONNCL,IONUCL]{Peter N Taylor\corref{jsa}}
\author[UFRJ]{Bruno Mota\corref{jsa}}

\cortext[mycorrespondingauthor]{Correspondence should be directed to: yujiang.wang@ncl.ac.uk}
\cortext[jsa]{Denotes equal contributions}
\address[ICOS]{Interdisciplinary Complex Systems Group, School of Computing, Newcastle University, Newcastle upon Tyne, UK}
\address[IONNCL]{Faculty of Medical Sciences, Newcastle University, Newcastle upon Tyne, UK }
\address[IONUCL]{UCL Queen Square Institute of Neurology, London, UK }
\address[Tuebingen]{Graduate Training Center of Neuroscience, University of T\"ubingen, T\"ubingen, Germany}
\address[CMIC]{Centre for Medical Image Computing (CMIC), University College London, London, UK}
\address[Dep]{Department of Clinical and Experimental Epilepsy, University College London, London, UK}
\address[QU]{Department of Medicine, Division of Neurology, Queen’s University, Kingston, Canada}
\address[Chalfont]{Epilepsy Society MRI Unit, Chalfont St Peter, UK}
\address[UFRJ]{Institute of Physics, Federal University of Rio de Janeiro, Brazil}

\begin{abstract}
Quantification of brain morphology has become an important cornerstone in understanding brain structure. Measures of cortical morphology such as thickness and surface area are frequently used to compare groups of subjects or characterise longitudinal changes. However, such measures are often treated as independent from each other.

A recently described scaling law, derived from a statistical physics model of cortical folding, demonstrates that there is a tight covariance between three commonly used cortical morphology measures: cortical thickness, total surface area, and exposed surface area. 

We show that assuming the independence of cortical morphology measures can hide features and potentially lead to misinterpretations. Using the scaling law, we account for the covariance between cortical morphology measures and derive novel independent measures of cortical morphology. By applying these new measures, we show that new information can be gained; in our example we show that distinct morphological alterations underlie healthy ageing compared to temporal lobe epilepsy, even on the coarse level of a whole hemisphere.

We thus provide a conceptual framework for characterising cortical morphology in a statistically valid and interpretable manner, based on theoretical reasoning about the shape of the cortex.

\end{abstract}


\end{frontmatter}


\section{Introduction}
Since magnetic resonance imaging has become widely available, the quantification of brain morphology has become a standard tool. Differences in brain morphology between a control and a comparator cohort are often reported for many processes in health and disease. Alterations in brain morphology, however,  may be non-specific; many processes appear to be associated with similar changes. For example, in healthy ageing, many studies report a thinning of the cortex as the predominant characteristic (e.g. \cite{bajaj_brain_2017,hutton_comparison_2009}). Similarly, many brain disorders (e.g. bipolar disorder \cite{hibar_cortical_2018}, schizophrenia \cite{van_erp_cortical_2018}, temporal lobe epilepsy \cite{whelan_structural_2018}, and Alzheimer’s disease \cite{dickerson_cortical_2009}) also feature cortical thinning as the predominant cortical alteration compared to controls. Such observations can lead to naïve conceptualisations, e.g. that the biological processes determining cortical thickness are particularly ``fragile'', or that certain brain disorders are the result of ``premature ageing''. In this study, we demonstrate that such concepts are inferences based on a univariate view of the brain morphology data. When considering a multivariate view, accounting for covariance, the alterations in different processes can be shown to be more specific and distinct.

One such multivariate view of brain morphology data has been proposed in the context of quantifying cortical folding. Based on a statistical physics model describing cortical folding, Mota \textit{et al.} predicts that cortical thickness $T$, cortical surface area $A_t$, and exposed surface area $A_e$ should be tightly linked by a scaling law $A_t \sqrt{T} = k A_e^{5/4}$, where $k$ is a constant. This equation has been derived based on the assumption that the cortex is a tissue of finite thickness that folds in a way that balances compressive mechanical forces (e.g. axonal tension in white matter and cerebral spinal fluid pressure \cite{franze_mechanical_2013,bayly_mechanical_2014}) with the imperative that it must be self-avoiding. The resulting scaling law has been confirmed by empirical data across mammalian species\cite{science2015}, individual humans\cite{pnas2016}, and even across different lobes of the same brain\cite{commbiol2019}. This scaling implies a tight covariance of the three morphological variables, whereby changes in one variable must be balanced by changes in the other variables. Conceptually, this means that, for example, if cortical thickness and total surface area are specified (by, e.g., the specifics of various neuroproliferative pathways during development), then its exposed area and volume follow as a physical consequence. More succinctly, cortical morphology variables are not independent of each other and and cannot to vary freely. 

The practical implication of the scaling law is that the three morphological quantities of cortical thickness, cortical surface area, and corresponding exposed surface area should not be treated independently when assessing brain morphology. Independent comparisons of these quantities may result in incorrect conclusions when not accounting for the covarying morphological features. For example, comparing cortical thickness between two groups without accounting for differences in surface area and exposed area (morphological covariates) would be as naive as comparing an Alzheimer's group against a control group without accounting for group differences in age (a biological covariate). 

Is there then a more systematic way of analysing cortical morphology that accounts for the covariance between morphological variables? The scaling law itself provides a natural way forward. In mathematical terms, the scaling law provides a ``principal component'' decomposition of the three morphological variables, and the resulting components are independent of each other and can be used to quantify cortical morphology. We will demonstrate this principle and show that brain disorders (temporal lobe epilepsy in our example) that \textit{appear} morphologically similar to ageing actually undergo distinct morphological changes to ageing.

\section{Methods}

\subsection{Data and Demographics}
To study the alterations associated with ageing, we used T1 and T2 weighted MRI brain scans from The Cambridge Centre for Ageing and Neuroscience (Cam-CAN) dataset (available at \url{http://www.mrc-cbu.cam.ac.uk/datasets/camcan/} \cite{shafto2014cambridge,taylor2017cambridge}). 
To study the alterations associated with temporal lobe epilepsy (TLE), we used the same subjects as in Taylor \textit{et al.} \cite{Taylor2018} and focused on the T1 weighted images. Cam-CAN used a 3T Siemens TIM Trio System with 1~mm isotropic voxel size (for more details see \cite{shafto2014cambridge,taylor2017cambridge}). The TLE dataset was obtained on a 3T GE Signa HDx scanner (General Electric, Waukesha, Milwaukee, WI) using a coronal T1-weighted volumetric acquisition with 170 contiguous 1.1~mm thick slices (matrix, $256 \times 256$; in-plane resolution, $0.9375 \times 0.9375$~mm), for more details see \cite{Taylor2018}. 

From the Cam-CAN dataset we retained 644 subjects that successfully completed preprocessing (recon-all - see next section) without errors. From these subjects we selected all subjects between 23 and 27 years old (inclusive) as our reference cohort, and all subjects between 33 and 37 (inclusive) as the comparison cohort. This resulted in 34 subjects in the reference cohort and 56 subjects in the comparison cohort. Note that in Supplementary Data, we show results for more groups from the Cam-CAN dataset to demonstrate robustness of the results. The TLE dataset included 53 patients with TLE (comparison cohort) and 30 controls (reference cohort). The control cohort spans an age range of 19-64 years, and the TLE cohort spans an age range of 19-67 years.

\subsection{Data processing}
The MR images of both datasets were first preprocessed by the FreeSurfer 6.0 pipeline \textit{recon-all}, which extracts the grey-white matter boundary as well as the pial surface. These boundaries were then quality checked and manually corrected where needed. Next, the relevant quantities (pial surface area, cortical thickness, and exposed surface area) were extracted from the FreeSurfer output files and assembled into one table (code is available in \cite{tools-repo}). In the following, the analysis is always hemisphere based, as in our previous work \cite{science2015,pnas2016}. We did not perform a more regionalised analysis, which is also possible \cite{commbiol2019}, as we wish to demonstrate the principle of independent morphological variables rather than describe the exact nature of morphological changes in a particular process. Future work using the principle demonstrated here may wish to include regionalised measures, as we discuss later.

\subsection{Scaling law analysis, and new morphological measures}


Throughout the paper, we use a log-space representation of all variables to allow us to combine variables linearly. We also chose variables that have all dimensions of area ($A_t,A_e$ and $T^2$) to allow an easier interpretation of the combination of variables. In this representation, the scaling law $\log A_t + \frac{1}{4}\log T^2 = \log k + \frac{5}{4} \log A_e$ defines the plane along which most cortices are situated. By isolating the parameter $k$, we can define a new vector $K=\log k=\log A_t - \frac{5}{4}\log A_e +\frac{1}{4}\log T^2$, perpendicular to this plane. We have previously hypothesised $K$ to be a combination of axonal tension and cerebral spinal fluid (CSF) hydrostatic pressure. We thus call $K$ the tension component. Note that $K$ is a constant for a homogeneous adult cohort of human subjects \cite{pnas2016}, and varies little across species \cite{science2015}.

It is remarkable that $k$ is a dimensionless quantity. This means that if two cortices are isometrically scaled versions of one another (i.e., same shape, different size), they will have the same $K$ value. Mathematically, isometric scaling means all areas, $A_t$, $A_e$, and $T^2$, are multiplied by a common numerical factor. This corresponds to movement in the direction $I=\log A_t + \log A_e + \log T^2$, the so-called isometric component, which is perpendicular to $K$. For a third and last element of our new set of orthogonal vectors, we use the cross-product of $K \times I = S=\frac{3}{2}\log A_t + \frac{3}{4}\log A_e -\frac{9}{4}\log T^2$, or shape factor, as the direction that is perpendicular to both $K$ and $I$. $K$ is determined by the scaling law and we chose the directions of $I$ based on the interpretability of the isometric component. $S$ is then simply the component that is perpendicular to both $K$ and $I$.

The parameter $I$ captures all the information about the size of the structure. Changing $I$, while keeping the other parameters constant, corresponds to isometrically shrinking or expanding a shape. One can think of the $I$ parameter, or coordinate, for any particular shape as a measure of size that carries no information about shape.

Conversely, the plane $K \times S$, henceforth called the isometric plane, carries only information about shape, and is not affected by size or changes in overall scaling. Any direction in this plane corresponds to the logarithm of a dimensionless parameter (mathematically, the sum of its vector coefficients is zero).


In our definition of the new components, we did not normalise the components (i.e. $K,S$ and $I$ have different length). This is because we standardise (z-score) all subjects relative to the reference group, and normalise the data in this way. However future application may want to use normalised vectors.

\subsection{Age and sex correction}
In order to investigate the effect of temporal lobe epilepsy alone, without the confounding effects of age and sex, we linearly regress out the effect of age and sex from all three log-transformed morphological variables cortical thickness, cortical surface area, and exposed surface area. We do this by deriving the linear regression coefficients from the control cohort, and applying them to both the control and the patient cohorts. Interaction between age and sex was not modelled.

To study the effect of ageing, we used two groups within a small age range (23-27 years old \textit{vs.} 33-37 years old). Thus, we did not perform the age correction, but only a sex correction by performing a mean centering for both sexes independently.

\subsection{Statistical analysis}
To statistically compare the effects of ageing and temporal lobe epilepsy, we standardise all quantities relative to the respective control cohort and report all effects in terms of effect sizes. This was achieved by converting all measures in all subjects to z-scores relative to the mean and standard deviation of the respective reference/control cohort. Hence, all quantities reported are in terms of z-scores. To measure the mean difference between the reference cohort and the comparison cohort (older ageing group or TLE cohort), we show the distribution of bootstrapped means (over 100 resampling iterations) of the z-scores for both groups as violin plots.  Note that as we are using a distribution of bootstrapped means, the mean of this distribution should be very close to zero in the reference groups, but may not be exactly zero in all cases due to the stochastic nature of bootstrapping.

To measure average effect between groups (termed $d$ in the following), we then form the difference between the average bootstrapped means (of reference \textit{vs.} comparison groups).  Note that $d$ is positive if the comparison group (older age group, or TLE) has a higher mean value than the reference group, and \textit{vice versa}.

As we were interested in group effects of ageing and TLE, we focused our attention only on the group mean estimation. The bootstrapping was a data-driven way to obtain a more representative mean group effect that was not driven by few outliers.

We also report p-values for statistical significance in the comparison of groups, only with the purpose to be consistent with previous studies, but not for subsequent use (e.g. to select features). All p-values are calculated using the Wilcoxon ranksum test on the raw data (i.e not the bootstrapped means).

\subsection{Data availability}
Code for extraction of raw cortical measures can be found on Zenodo \cite{yujiang_wang_analysis_2019} and Github: \url{https://github.com/cnnp-lab/CorticalFoldingAnalysisTools}.

Data underlying the figures in this paper and the corresponding code can be found on Github: \url{https://github.com/cnnp-lab/2020Wang_TLEFoldingHemi}

\section{Results}

\subsection{Morphological changes in TLE appear to be the same as in ageing}

\begin{figure}[h!]
  \centering
  \hspace*{-1.5cm}
  \includegraphics[width=140mm]{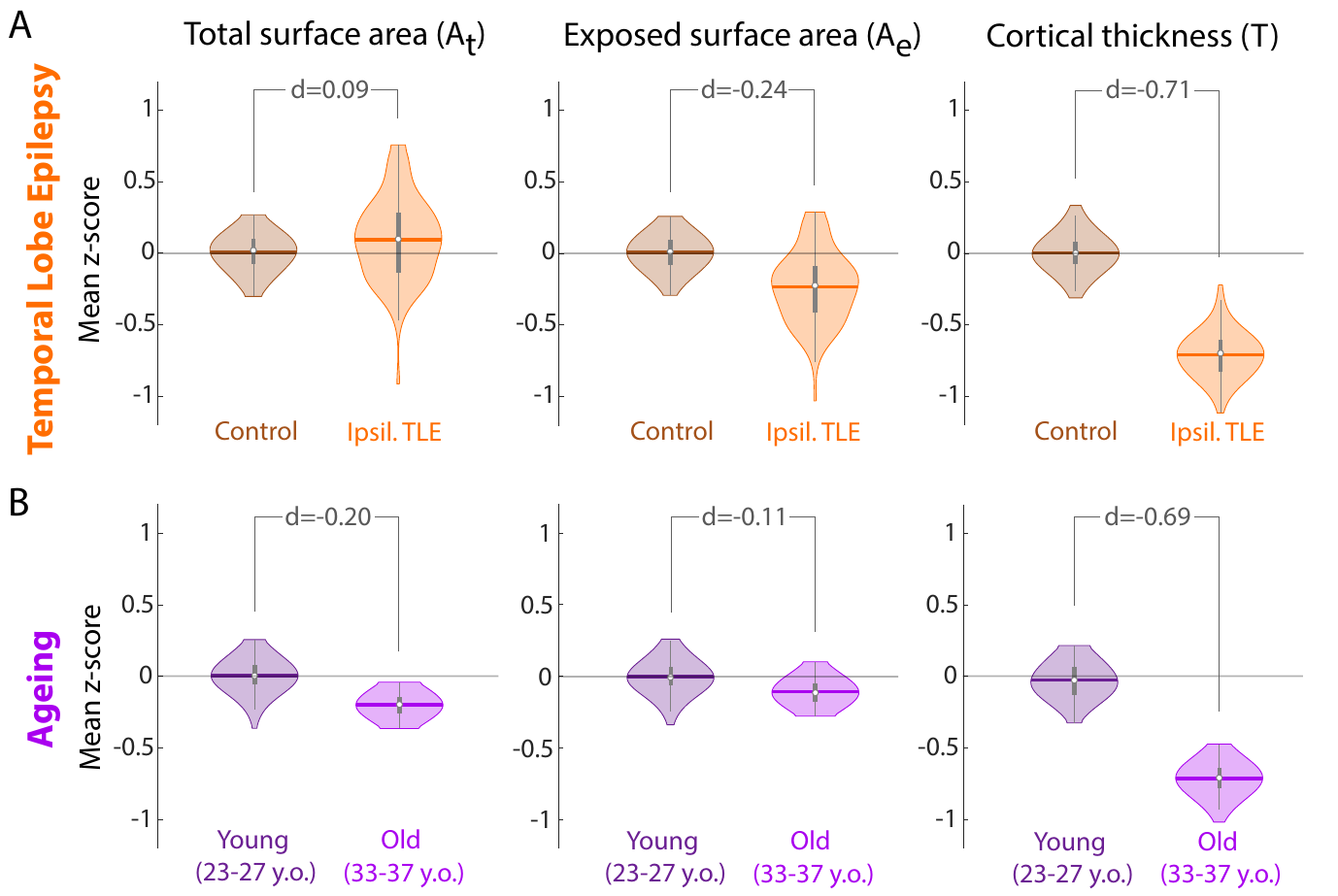}
  \caption{\textbf{Morphology changes in TLE appear similar to those in healthy ageing.}  \textbf{(A)} Morphology changes in TLE in the ipsilateral hemisphere compared to a control cohort measured as z-scores. Violin plots show the distribution of bootstrapped mean z-scores. Age and sex correction was performed before the comparison. \textbf{(B)} Morphology changes in healthy ageing comparing a younger and older group of adult subjects, measured as z-scores relative to the younger subject group. Violin plots show the distribution of bootstrapped mean z-scores. Sex correction was performed before the comparison. (A \& B) All morphological measures are in terms of a whole cortical hemisphere and log-scaled before analysis. Each hemisphere was treated as a separate datapoint. Beeswarm plots with raw data points are presented in Supplementary Data.}
  \label{Fig1_OrigVar}
\end{figure}

In many diseases, average cortical thickness is the most consistently decreasing variable relative to controls. Temporal lobe epilepsy (TLE) is no exception. In our data (Fig.~\ref{Fig1_OrigVar}),  average cortical thickness of the entire ipsilateral hemisphere (cortical ribbon) is substantially reduced in patients relative to controls ($d=-0.71$, $p=0.0008$). Total and exposed surface area do not appear substantially altered ($|d|<0.3$, $p>0.05$).

The same patterns of alteration are observed in the healthy ageing process. In our cross-sectional data, cortical thickness is again substantially  reduced in older subjects ($d=-0.69$, $p=0.00004$), while surface areas remain relatively unaltered.

Given these parallel alterations in brain morphology, one might be tempted to speculate that an accelerated ageing process underlies morphology alterations in TLE (and other diseases showing the same pattern). However, we will demonstrate in the following section that this would be an erroneous conclusion based on raw and, in this case, less ``informative'' measures of brain morphology,  analysed in a univariate manner, neglecting the covariance between these measures.

\subsection{The universal scaling law describes covariance of raw morphology measures}

\begin{figure}[h!]
  \centering
  \hspace*{-1.5cm}
  \includegraphics[width=140mm]{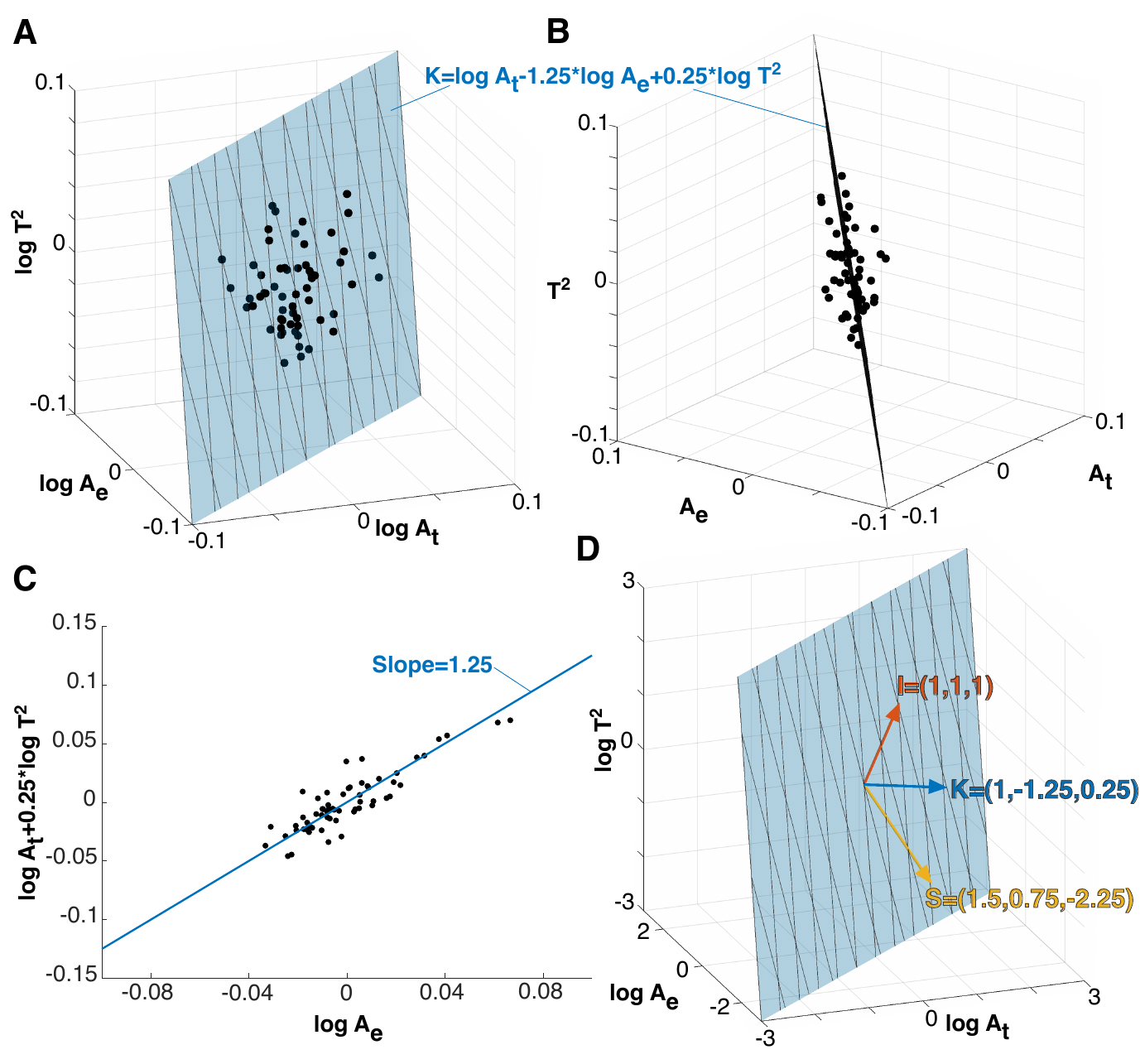}
  \caption{\textbf{Universal scaling law describes the covariance of the raw morphological measures.}  \textbf{(A)} Three raw morphoplogy measures span a 3D space, where each cortex is a data point (black dots). Here we used the control group in the TLE dataset as an example for the purpose of illustration. The data points roughly lie on the plane described by the universal scaling law (blue plane). \textbf{(B)} different viewing angle of the same data shown in (A). \textbf{(C)} Projection of data into a 2D space, which was previously used to visualise the scaling law. The blue line now represents the projected plane from (A) and (B). \textbf{(D)} 3D view of scaling law plane and viewing angle as in (A). The normal vector of the scaling law plane (K) is shown as a blue vector. Two perpendicular vectors (S and I) can be defined, and together they span the 3D space.  All morphological variables are logscaled and age corrected in this figure. Cortical thickness is presented as thickness squared so that the 3D space has units of area in all dimensions.}
  \label{Fig2_UniversalLaw}
\end{figure}

Any given cortex can be represented as a point in the $\log A_t \times \log A_e \times \log T^2$ space (Fig.~\ref{Fig2_UniversalLaw}A,B), which has units of area in all dimensions. By plotting the TLE control cohort in this way, it is evident that the raw morphological measures $A_t$, $A_e$, and $T$ covary tightly in this space (Fig.~\ref{Fig2_UniversalLaw}. When superimposing the plane described by the scaling law ($\log A_t + \frac{1}{4}\log T^2 = \alpha \log A_e + \log k$, where $\alpha$ is theoretically predicted to be $\frac{5}{4}$), we can see that it fits well to describe the covariance of the raw morphological measures (Fig.~\ref{Fig2_UniversalLaw}A,B,C).  Both the TLE control, as well as the patient group follow this scaling law ($\alpha$ slope 95\% CI 1.1548 - 1.4260 and 0.9665 - 1.2827, respectively), also see Fig.~\ref{Fig2_UniversalLaw}C. Note that because of age-correction, all controls line up on the plane described by $\log k = K =0$.

In other words, the scaling law provides a decomposition of the raw morphological measures: The normal vector to the plane is $(1,-1.25,0.25)$ (Fig.~\ref{Fig2_UniversalLaw}D) where the first, second, and third dimensions are $A_t,A_e$ and $T^2$, respectively. By calculating $K= \log A_t - 1.25 \log A_e +0.25 T^2$ we can obtain a value for $K$ for every cortex from their raw morphological measures $A_e, A_t$ and $T$. Based on our model \cite{science2015}, $K$ can be interpreted as a combination of axonal tension and CSF hydrostatic pressure \cite{pnas2016,commbiol2019}, we thus call $K$ the tension term.

The vector $(1,1,1)$ describing isometric scaling (i.e. changing $A_t,A_e$ and $T^2$ by the same proportion, which is equivalent to stretching/shrinking the brain in all direction equally) is perpendicular to the previous normal vector. We choose this to be the second component as it has a direct interpretation, and it is also independent of $K$ in our dataset (Pearson's $\rho=0.09$, $p=0.45$). Again, it can be calculated as $I= \log A_t + \log A_e + \log T^2$ (isometric term) from the raw morphological variables. It can be understood to carry information about the size of the cortex only, without containing any information about shape.

The third perpendicular vector that is the cross-product of the two previous ones is $(\frac{3}{2},\frac{3}{4},-\frac{9}{4})$. We will call this the shape term $S=\frac{3}{2} \log A_t + \frac{3}{4} \log A_e -\frac{9}{4} \log T^2$. Again, $K$ and $S$ are independent (Pearson's $\rho=0.02$, $p=0.85$). While $I$ only carries information about size, $K$ and $S$ only carry information about shape. This also means that for the same $K$ (which is the case for all healthy human adults of the same age \cite{pnas2016}), $S$ is the only term that describes any changes in shape.

Here, the choice of $I$ and $S$ did not follow a data-driven principal component analysis. Instead, we choose directions predicted by the scaling law and that are interpretable, as our intention is to provide an illustrative demonstration of a set of new independent morphological variables. 

\subsection{The universal scaling law defines a new set of independent morphological measures}

\begin{figure}[h!]
  \centering
  \hspace*{-1.5cm}
  \includegraphics[width=140mm]{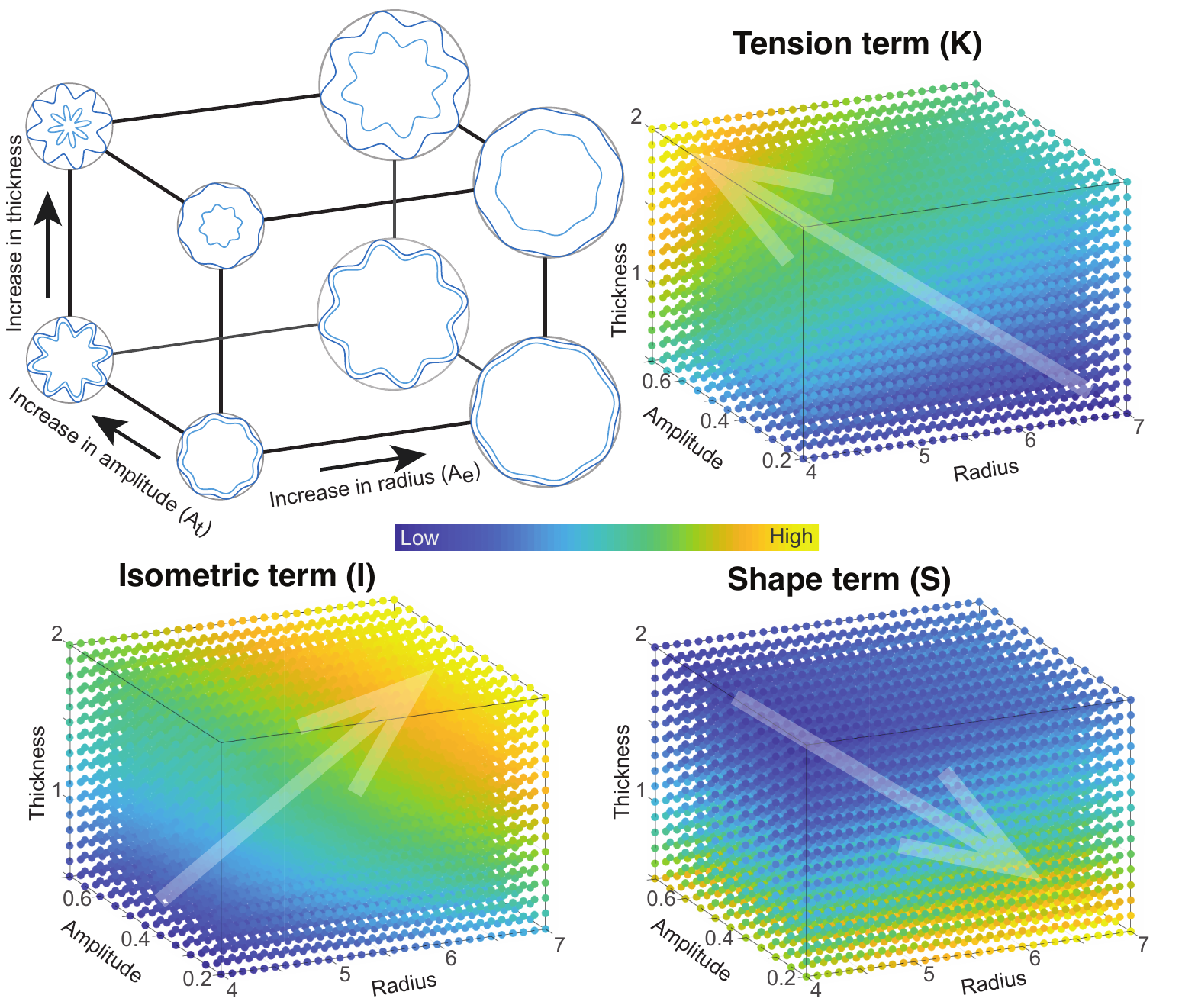}
  \caption{\textbf{Schematic to provide intuition for the three projection terms K, S, and I}  Simulations of basic folded ribbons as sinusoidal oscillations on a circle. In this shape we can change the overall radius of the encapsulating circle ($A_e$), the thickness ($T$) encapsulated by the outer and inner oscillations (dark and light blue), and the amplitude of the oscillations, which dictate the total length of the oscillation ($A_t$). By scanning the radius, thickness, and oscillation amplitude in a 3D space, we can calculate the corresponding value for the $K$, $S$, and $I$ term at different points in this space (colour map). Transparent arrows point in the directions of change of $K$, $S$, and $I$.  Through visualising the changes in $K$, $S$, and $I$ in this 3D space, we provide an intuition for how the three terms relate to parameters in a simple folded structure.}
  \label{Fig3_KISIntuition}
\end{figure}

To provide an intuitive understanding of the new morphological variables, we provide a schematic illustration of the variables on a 2D shape (a circular sinusoidal ribbon) in Fig.~\ref{Fig3_KISIntuition}. We can intuitively parametrise this shape with the overall radius of the circle, the amplitude of the sinusoid, and the thickness of the ribbon to describe changes in $A_e,A_t$ and $T$ respectively. We then demonstrate how such changes map onto the new morphological variables $K,S$ and $I$. $I$ corresponds to a measure of size, as expected. It increases with the thickness of the ribbon as well as the overall circle radius, for a constant sinusoid amplitude. $S$ increases with a combination of overall radius and amplitude of the sinusoid, but decreases with thickness of the ribbon. Finally, $K$ increases with thickness and sinusoid amplitude, but decreases with overall radius.

\subsection{The scaling law derived morphological measures show differences between TLE and ageing}

\begin{figure}[h!]
  \centering
  \hspace*{-1.5cm}
  \includegraphics[width=140mm]{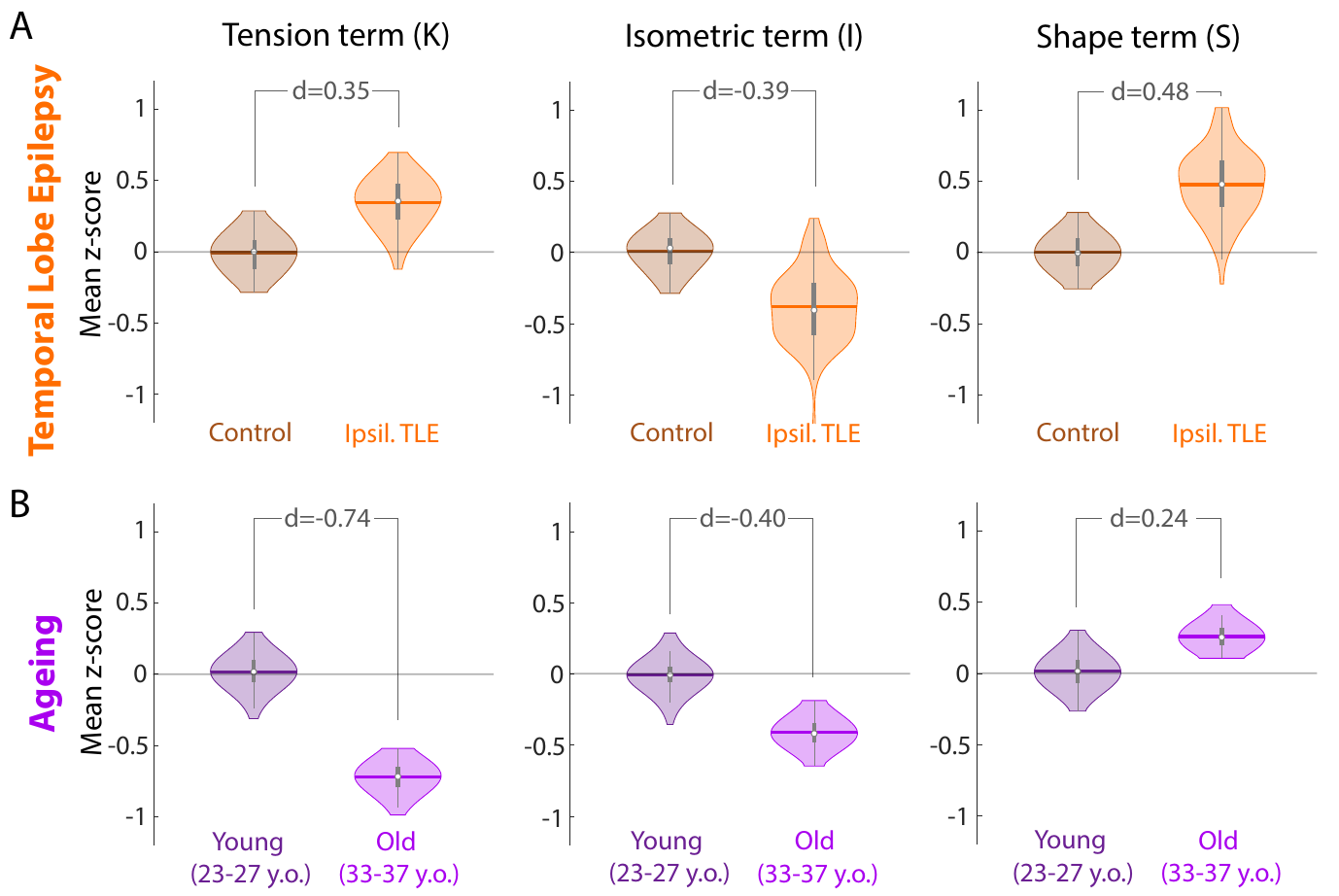}
    \caption{\textbf{Morphological changes in K differ in TLE compared to healthy ageing.}  \textbf{(A)} Morphological changes in $K$, $S$, and $I$ in the ipsilateral hemisphere in TLE compared to a control cohort measured as z-scores relative to controls. Violin plots show the distribution of bootstrapped mean z-scores. Age and sex correction of original morphological measures was performed before the comparison. \textbf{(B)} Morphological changes in healthy ageing comparing a younger and older group of adult subjects, measured as z-scores relative to the younger subject group. Violin plots show the distribution of bootstrapped mean z-scores. Sex correction of original morphological measures was performed before the comparison. (A \& B) All morphological measures are in terms of a whole cortical hemisphere. Each hemisphere was treated as a separate datapoint. Beeswarm plots with raw data points are presented in Supplementary Data.}
  \label{Fig4_KISData}
\end{figure}

Equipped with the new measures $K$, $S$, and $I$, we now re-examine our initial observation that the morphological alterations in TLE resemble those of ageing. Fig.~\ref{Fig4_KISData} shows that both ageing, and TLE are associated with similar alterations in $S$ and $I$ with a similar effect size (decrease in $I$ with $d \approx -0.4$, and an increase in $S$ with $d =0.24$ and $d=0.48$). However, TLE is associated with an increase in $K$ compared to controls ($d=0.35$), whereas ageing is clearly associated with a strong decrease in $K$ ($d=-0.74$). In other words, in terms of the tension term, brain morphological changes in TLE differ from those in healthy ageing.

\subsection{Outlook: Trajectories of disease and ageing processes}
We can additionally visualise the average effects from Fig.~\ref{Fig4_KISData} as datapoints in the three-dimensional space of spanned by $K \times S \times I$, in terms of effect sizes in each of those three independent variables. In other words, each process/condition (ageing, TLE) can be understood as an alteration in $K$, $S$, and $I$ relative to controls/reference. By placing the reference at the origin of this space (0,0,0), one can visualise the effect of each process a datapoint corresponding to their effect in $K,S$ and $I$.

\begin{figure}[h!]
  \centering
  \includegraphics[width=90mm]{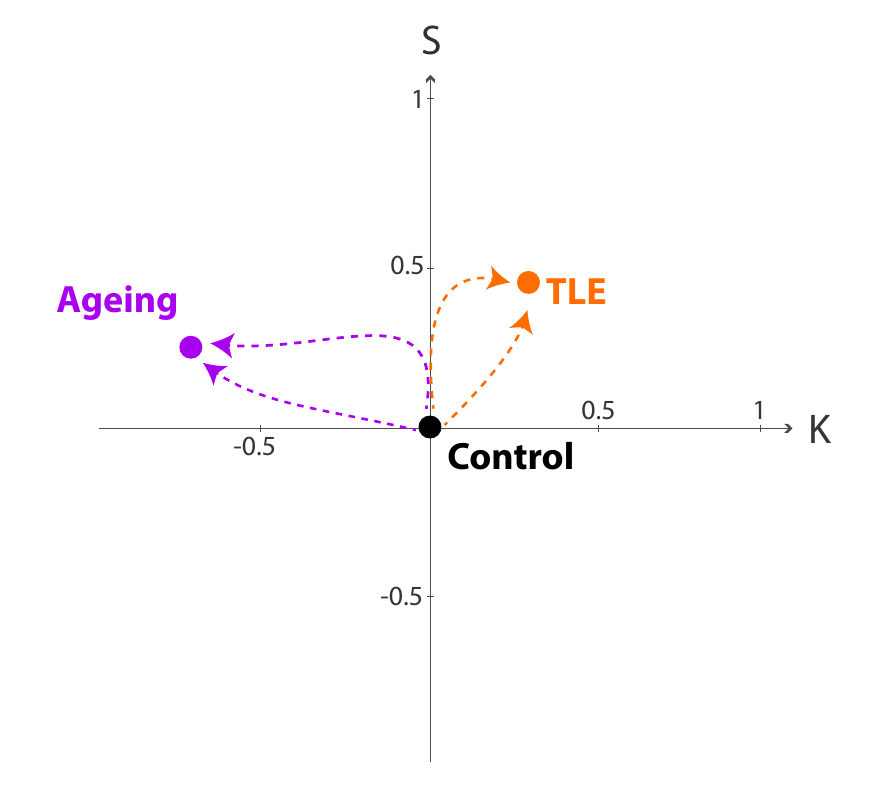}
    \caption{\textbf{Trajectories of morphology changes in health and disease.} Visualising the changes in ageing and TLE (same data as Fig.~\ref{Fig4_KISData}) in the 2D projection into $K$ and $S$ as trajectories from the origin. We chose to show a 2D projection of $K \times S \times I$ space for simplicity. Both ageing and TLE process have been centered according to their respective control group. The respective datapoints are derived from the corresponding $d$ values in each component from Fig.~\ref{Fig4_KISData}. Dashed lines indicate possible (hypothesised) trajectories. Note that trajectories can in theory move in any direction in this space, as the axes are now independent. Shared trajectories would reflect true shared mechanisms of brain morphology change.}
  \label{Fig5_KIS3D}
\end{figure}

Fig.~\ref{Fig5_KIS3D} shows the control/reference as a point at the origin. Ageing and TLE are represented as two separate datapoints in this space, and clearly separated by the $K$ component. In such a representation it becomes clear that both processes/conditions must have followed a trajectory (indicated by dashed lines in Fig.~\ref{Fig5_KIS3D}) that links the control condition with the disease or ageing ``end points''. These trajectories could in theory follow any path, and are not restricted to particular parts of the space, as the variables are independent. The conceptual advance of this paper is to enable such a space where the axes are independent. This now allows for an unbiased study of disease trajectories \cite{jensen_temporal_2014} on an individual, or group level. Clustering of trajectories now will reflect shared disease mechanisms, rather than unaccounted covariance between variables.

\section{Discussion}
Using TLE and ageing as examples, we show limitations of using and interpreting  morphological measures in an univariate manner. To account for the existing covariance between morphological variables, we suggest using new independent variables/measures. These independent measures clearly demonstrate that although TLE appears morphologically similar to the ageing process on the surface, these two processes are in fact distinct in terms of their morphological alterations.

Although we used a whole-hemisphere approach in our examples, parallel arguments also hold for region-specific changes. We used the whole-hemisphere analyses to demonstrate the principle that covarying morphological measures need to be accounted for, however, we acknowledge that ageing and disease processes should not necessarily be simplified to a whole-hemisphere view when trying to understand their biological mechanisms. Further, some conditions may only have localised effects. Thus, to apply our principle to practical questions of biological mechanisms, we are now extending our work to local/regional measures of brain morphology.  Indeed, we previously showed that the scaling law also holds for lobes/areas of the same brain\cite{commbiol2019}. This means that local measures of $K,S$ and $I$ can be derived, based on local measures of cortical thickness, total and exposed surface area. One approach is to use a parcellation of the brain. However, in our previous work we noted that information about the local geometry of the cortex is also required, and a fine-grained parcellation is therefore not recommended. Alternatively, as discussed in our previous work, it is in theory possible to derive a point-wise/voxel-wise estimate of $K,S$ and $I$ on the cortical surface, following the same principle as the derivation of the local gyrification index \cite{schaer_surface-based_2008}. This point-wise estimates may help discovery of covert local abnormalities in focal epilepsies, or identification of local neurodegeneration to act as a sensitive biomarker for effects of disease modifying drugs, in e.g. Multiple Sclerosis.  To make this extension of our work possible and accessible to others, we have also made our MATLAB code available (\url{https://github.com/cnnp-lab/CorticalFoldingAnalysisTools}), including the processing of regionalised measures.

TLE and ageing differ most in the tension term $K$ in our analysis. While ageing is associated with a decrease in $K$ (in agreement with previous work \cite{pnas2016,commbiol2019}), TLE is associated with an increase in $K$ compared to controls. In the theoretical derivation of the scaling law, $K$ appears as a term that describes the pressure applied to the cortical surface \cite{science2015}. We have previously interpreted this as CSF pressure, or the average white matter tension \cite{essen_tension-based_1997,xu_axons_2010,franze_mechanical_2013,bayly_mechanical_2014} exerted on the grey matter. We previously speculated that the decrease in $K$ could be interpreted as a slackening of white matter tension, which may be in agreement with magnetic resonance elastography findings in ageing \cite{Arani2015,Hiscox2018}. By the same argument, an increase in $K$ in TLE could be interpreted as a stiffening of the white matter. However, there are no studies in the field yet to confirm or reject this hypothesis. Further evaluation of the biological basis of these findings is necessary, with consideration of diffusion properties in white and grey matter and measures of stiffness, for example as measured with MR elastography \cite{murphy_mr_2019}.

In the scaling law, we consider the variables of cortical thickness, cortical surface area, and exposed surface area. Other morphological variables such as cortical volume or intracranial volume are additional frequently-used quantities. In particular, intracranial volume is often used as a covariate to account for ``brain size''. 
In our simplified theoretical derivations, cortical volume and intracranial volume are assumed to be directly determined by a combination of thickness, total surface area and exposed surface area. However, these volume variables may well hold additional information not captured in the scaling law (see e.g. \cite{Wierenga2014}), and we hypothesise that such variables may explain the residual variance in our data. Future work could perform a data-driven principal component decomposition in a well-controlled cohort of subjects and across healthy development.

Our work has some conceptual parallels and distinctions of our work with a few established neuroimaging analysis approaches. One prominent approach that is also concerned with the covariance of morphological quantities is the so-called ``structural covariance analysis'' \cite{AlexanderBloch2013}. In that approach, the covariance is measured between different regions of the brain in terms of one morphological measure (e.g. cortical thickness), essentially assessing which regions change together across subjects. The popular approach is to then understand the covariance as a matrix that describes a network, and compare these networks between groups. The independent variables $K,S$ and $I$ may be more advantageous in terms of its reliability and comparability \cite{carmon_reliability_2019} for structural covariance analysis.  
A related approach has been termed ``morphometric similarity'', where for a single subject, the covariance between brain region is derived based on their similarity across a range of morphological measures \cite{Seidlitz2018}. However, note that both types of approaches are concerned with covariance between brain regions, rather than covariance between morphological measures. We envisage that a comprehensive framework for cortical morphology would encompass both aspects in the future.

Apart from covariance of morphological quantities, our work is also related to measures of ``fractal dimension'' of the brain shape (see e.g. \cite{madan_cortical_2016}). Indeed, a natural way in which such a universal scaling law could arise would be if cortices were self-similar (in a statistical sense) down to some fundamental length scale proportional to cortical thickness, approximating a fractal with fractal dimension 5/2 (see \cite{pnas2016,commbiol2019}). This is in the same range as recent reports of the empirically measured fractal dimension \cite{madan_cortical_2016,madan_testretest_2017}. However, this is just an indication, not proof, of the hypothesized self-similarity. Future studies will have to demonstrate that the brain actually approximates a fractal object, by e.g. relating the scaling of a single cortex undergoing an process of iterated coarse-graining versus the scaling of different cortices. If confirmed, then additional concepts from fractal geometry could further enhance our analysis and understanding of the brain's folded shape.

Finally, we proposed the notion of ``trajectories'' in morphological space (spanned by independent variables), building on related previous work (e.g. \cite{young_uncovering_2018}). A key implication of such trajectories is that different brain processes (or disorders) may cluster in terms of their trajectories, or share parts of their trajectories, potentially indicating shared drivers/pathways/modulations \cite{jensen_temporal_2014,taylor_early_2020}. Especially with a comprehensive region-specific and cross-region analysis of cortical morphology we expect clusters of directions to emerge. On an individual subject level, our approach may also help to develop more sensitive and specific biomarkers. Moreover, current efforts to relate morphological alterations to genetic alterations (e.g. \cite{Seidlitz2019}) may help to develop an atlas of principal trajectories, and shed light on potential corresponding biological mechanisms.

In summary, our work represents a significant conceptual advance by contributing independent cortical morphology measures that can be interpreted without being hampered by other unaccounted morphological covariates. Using these independent measures we demonstrated that temporal lobe epilepsy, which appeared to resemble premature ageing in terms of cortical morphology, is in fact characterised by distinct morphological changes from ageing. The same principle may resolve some of the existing confusion in the literature regarding morphology alteration in other brain conditions and processes. In future, we hope that systematic studies of brain morphology can be associated with the underpinning biological mechanisms, and become a useful tool in biomarker development and understanding the brain in health and disease.

\section{Acknowledgements}
We are grateful to the Epilepsy Society for supporting the Epilepsy Society MRI scanner. This research was supported by the National Institute for Health Research University College London Hospitals Biomedical Research Centre. 

Part of the data for this research was provided by the Cambridge Centre for Ageing and Neuroscience (CamCAN). CamCAN funding was provided by the UK Biotechnology and Biological Sciences Research Council (grant number BB/H008217/1), together with support from the UK Medical Research Council and University of Cambridge, UK.
 
We thank members of the CNNP lab (\url{www.cnnp-lab.com}) for discussions on the analysis and manuscript. PNT and YW gratefully acknowledge funding from Wellcome Trust (208940/Z/17/Z and 210109/Z/18/Z). JHN was supported by the Reece Foundation. GPW was funded by the MRC (G0802012, MR/M00841X/1). BM is supported by Funda\c{c}\~ao Serrapilheira Institute (grant Serra-1709-16981) and CNPq (PQ 2017 312837/2017-8). The authors declare no conflict of interest. The funders played no role in the design of the study.

\section*{Author contributions}
YW, PNT, and BM conceived the idea. 
YW wrote the code, performed all the analysis, and produced all the figures. TL validated the code.
GW, SBV, JdT, and JSD contributed the TLE data.
PNT processed the TLE data.
YW, TL, BL, JHN, and PNT inspected the Freesurfer processing of the data and performed manual corrections where needed. 
YW, PNT, and BM drafted the manuscript. 
All authors participated in critically reviewing and revising the manuscript.

\section*{Supplementary Data}
For supplementary data, please see the Github repository underlying the analyses of this paper: \url{https://github.com/cnnp-lab/2020Wang_TLEFoldingHemi}. In the subfolder \textit{figs} additional plots equivalent to Fig.~\ref{Fig1_OrigVar} and \ref{Fig4_KISData} can be found for two further age category comparisons. In the subfolder \textit{figs\_beeswarm}, all the equivalent figures using the raw datapoints can be found as beeswarm plots.

\bibliography{references.bib}

\end{document}